\documentclass[twocolumn,showpacs,preprintnumbers]{revtex4}%
\usepackage{amsfonts}
\usepackage{amsmath,epsfig}
\usepackage{graphicx}
\usepackage{dcolumn}
\usepackage{bm}
\usepackage{amssymb}
\usepackage{amsmath}%
\begin{document}
\title{Quark deconfinement and the duration of short Gamma Ray Bursts}
\author{Alessandro Drago$^{\text{(a)}}$, Andrea Lavagno$^{\text{(b)}}$,
Brian Metzger$^{\text{(c)}}$ and Giuseppe Pagliara$^{\text{(a)}}$}
\affiliation{$^{\text{(a)}}$Dip.~di Fisica e Scienze della Terra dell'Universit\`a di Ferrara and INFN
Sez.~di
Ferrara, Via Saragat 1, I-44100 Ferrara, Italy}
\affiliation{$^{\text{(b)}}$ Department of Applied Science and Technology, Politecnico
di Torino, Italy
Istituto Nazionale di Fisica Nucleare (INFN), Sezione di Torino, Italy}
\affiliation{$^{\text{(c)}}$ Columbia Astrophysics Laboratory, Columbia University,
New York, NY, 10027, USA}

\begin{abstract}
We propose a model for short duration gamma-ray bursts (sGRBs) based
on the formation of a quark star after the merger of two neutron
stars.  We assume that the sGRB central engine is a proto-magnetar,
which has been previously invoked to explain the plateau-like X-ray
emission observed following both long and short GRBs.  Here, we show
that: i) a few milliseconds after the merger it is possible to form a
stable and massive star made in part of quarks; ii) during the early
cooling phase of the incompletely formed quark star, the flux of
baryons ablated from the surface by neutrinos is large and it does not
allow the outflow to achieve a bulk Lorentz factor high enough to
produce a GRB; iii) after the quark burning front reaches the stellar
surface, baryon ablation ceases and the jet becomes too baryon poor to
produce a GRB; iv) however, between these two phases a GRB can be
produced over the finite timescale required for the baryon pollution
to cease; a characteristic timescale of the order of $\sim 0.1 $ s
naturally results from the time the conversion front needs to cover
the distance between the rotational pole and the latitude of the last
closed magnetic field line; v) we predict a correlation between the
luminosity of the sGRB and its duration, consistent with the data; vi)
our model also predicts a delay of the order of ten seconds between
the time of the merger event and the sGRB, allowing for the
possibility of precursor emission and implying that the jet will
encounter the dense cocoon formed immediately after the merger.
\end{abstract}

\maketitle

Both long duration (lGRBs) and short duration Gamma Ray Bursts (sGRBs)
start with a violent ``prompt" emission phase, which generally lasts a few tens
of seconds in the case of lGRBs and a few tenths of a second in sGRBs.
The prompt emission is in many cases followed by some form of
prolonged engine activity, commonly referred to as the
``Quasi-Plateau" (QP) in the case of lGRBs and ``Extended Emission"
(EE) in the case of sGRBs \cite{Norris&Bonnel06}. Beyond similarities
in their light curve behavior, sGRBs and lGRBs show remarkably similar
spectral properties \cite{Ghirlanda:2009de}.
This led to the suggestion that a similar central engine is acting
in both classes of GRBs, a sGRB being similar to a lGRB cut after $0.3(1+z)$ s \cite{Calderone:2014daa}.

The progenitors of lGRBs and sGRBs, on the other hand, are believed to
be quite different: the collapse of a massive star for long bursts
\cite{Woosley93} and the merger of two neutron stars (or of a neutron
star and a black hole) for the short bursts \cite{Paczynski86}.
In their original forms, both models postulated a hyper-accreting
black hole as the source of the relativistic outflow powering the GRB.
However, following the discovery of the prolonged emission,
a new model for the engine has grown in popularity, based on the
relativistic wind of a newly formed, rapidly rotating proto-magnetar
\cite{Thompson+04,Metzger:2010pp}.  The model was initially proposed
to explain the structure of lGRBs, but more recently it has been
adapted to interpret also sGRBs
\cite{Metzger+08,Bucciantini:2011kx,Rowlinson:2013ue} \footnote{Models
  for sGRBs based on the formation of a black hole are still actively
  discussed, see e.g.  Refs.\cite{Ciolfi:2014yla,Rezzolla:2014nva} and
  the criticism raised in \cite{Margalit:2015qza}.}.

GRB prompt emission results from dissipation within a relativistic jet
composed of electron-positron pairs, photons and a small (but
non-negligible) fraction of baryons \cite{Shemi&Piran90}. The latter plays a fundamental role by setting
the bulk Lorentz factor $\Gamma$ of the jet, with values of $\Gamma
\sim 10^2-10^3$ required to match the observational data in most jet
emission models \cite{Hascoet+14}.  In the case of a proto-magnetar, the requisite
baryon loading is set naturally by the rate of mass ablation from the
surface by neutrino heating \cite{Metzger:2010pp}.  The
duration of the initial prompt phase is therefore closely connected
with the cooling time of the proto-neutron star, which indeed
typically lasts tens of seconds or longer.  The subsequent
quasi-plateau is also powered by the still rapidly rotating magnetar,
but the emission properties are likely to change once the wind reaches a high
magnetization (pulsar-like) state after baryon loading ceases. Model fits
of QP light curve to the dipole spin-down luminosity
successfully describe the data \cite{Lyons:2009ka,Dall'Osso:2010ah}.
The same modeling applied to the EE of sGRBs
\cite{Rowlinson:2013ue} generally finds acceptable fits for similar
values of the initial rotation period $P \sim$ few milliseconds, but
the required dipole magnetic field strength $B$ is roughly an order of
magnitude larger than for lGRBs.

If the magnetar model is correct, a crucial question naturally
arises: what is the origin of the prompt emission for sGRBs?  If
broadly similar values of $P$ and $B$ are needed to describe the QP
and the EE, then why is sGRB prompt emission typically two orders of
magnitude shorter than in lGRBs?  The cleaner environment for the jet
to escape, and the larger peak temperature of the proto-magnetar (reaching
$\approx$ 50 MeV \cite{Sekiguchi:2011mc}) in NS mergers compared to
core collapse, would on the contrary suggest that the sGRB prompt
emission should last even longer than that of lGRBs!

In this Letter, we propose that due to the large mass of the
proto-magnetar formed after a neutron star merger its nature is that
of a quark star and not of a neutron star
\cite{Kurkela:2009gj,Drago:2013fsa} following the "two-families"
scenario of Ref.\cite{Drago:2013fsa,Drago:2015cea,Drago:2015dea} in
which light compact stars are made of hadrons while the most massive
ones are quark stars.  Quark stars are self-bound objects, such that
neutrinos with energies of a few tens MeV are not energetic enough to
ablate material from the surface of the star
\cite{Haensel:1991um,Dai:1998bb}. Therefore, after the complete
transformation of the newly formed compact star into a quark star, no
baryonic material can be ablated from its surface and the prompt
emission has rapidly to terminate.  We associate this brief phase of
cessation of the baryonic pollution with the duration of the prompt
emission in sGRBs.

Below, we will show that: 1) the formation of quark matter can take
place within a few milliseconds after the merger, stabilizing the
massive compact star; 2) the rate of baryon ablation from the surface
during the formation of the quark star (until its conversation is
complete) is too high to produce prompt GRB emission; 3) the duration
of the prompt emission in sGRBs can therefore be linked to the
switch-off of the baryonic emission, a process which we will show is
indeed expected to last a few tenths of a second. In this
way the prompt phase of sGRBs will look like that of lGRBs but cut at the moment of the 
switch-off, satisfying the analysis of Ref.~\cite{Calderone:2014daa}.

\begin{figure}[ptb]
\vskip 0.5cm
\begin{centering}
\epsfig{file=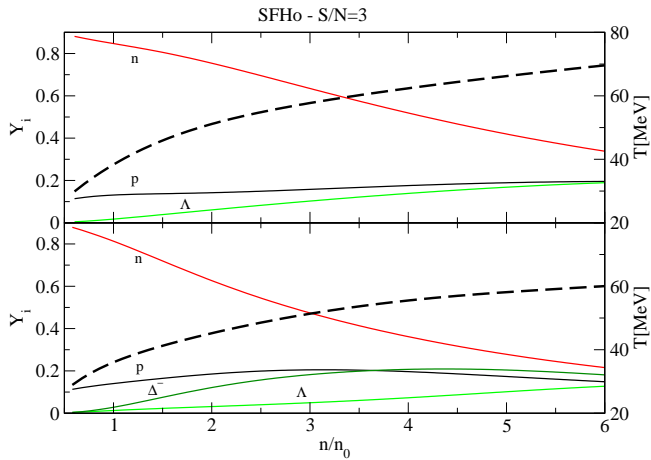,height=8.5cm,width=6cm,angle=-90}
\caption{Fractions of neutrons, protons, Lambda (and $\Delta$-resonances in the lower panel) as a function
of the density (left scale). Temperature (right scale). 
They are computed for matter having entropy per baryon of $S/N=3$. 
}
\label{fig:1}
\end{centering}
\end{figure}

We start by showing that, in the newly-formed compact star created by
the merger, the conditions for initiating quark deconfinement are
fulfilled. In Fig.~\ref{fig:1} we display the composition of matter at
beta-equilibrium and with an entropy-per-baryon $S/N=3$. This
corresponds to a temperature in the center of the merger
remnant of about 50 MeV, similar to what found by the simulations of
Ref.\cite{Sekiguchi:2011mc}. We have employed the EoS SFHo obtained in
\cite{Steiner:2012rk} which satisfies all existing constraints below
nuclear matter saturation density $n_0$ and we have taken into account the
possible formation of $\Delta$-resonances \cite{Drago:2014oja}. We
also show in Fig.1 the EoS excluding $\Delta$'s to prove that the
mechanism we are describing does not depend on the details of the
hadronic EoS. Importantly, note that hyperons are present already at
densities of the order of $n_0$ (in agreement with
Ref.\cite{Sekiguchi:2011mc}) due to the high temperature of the
system. Bubbles of deconfined quark matter (here described by the EoS
of Ref.\cite{Weissenborn:2011qu}) will start appearing throughout the
central region of the star on the time-scale of strong interaction
(the temperature is large enough that thermal nucleation can take
place \cite{DiToro:2006pq}) and will rapidly expand following the
scheme of Ref.\cite{Drago:2015fpa}. The central region will
deconfine on a time-scale of $\sim 3-4$ ms
\cite{Herzog:2011sn,Pagliara:2013tza} since in this initial phase the burning front
is strongly accelerated by hydrodynamical instabilities.

\begin{figure}[ptb]
\vskip 0.5cm
\begin{centering}
\epsfig{file=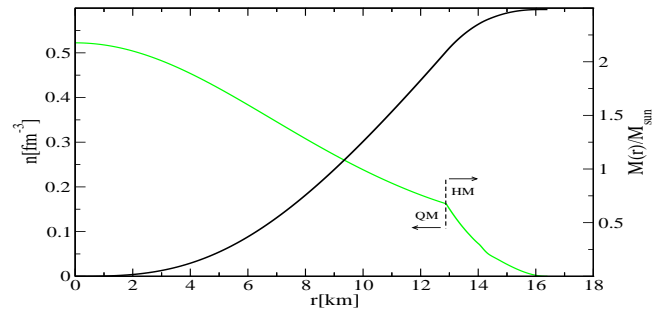,height=8.5cm,width=4cm,angle=-90}
\caption{Density profile (green line) and mass enclosed (black line) of the "hybrid" star formed after the rapid combustion as a function of the distance from the center.}
\label{fig:2}
\end{centering}
\end{figure}

This phase of rapid burning halts at a depth of a few kilometers below
the stellar surface, leaving the external layers unburnt and producing
in a few ms an intermediate configuration which is mechanically
stable, but not yet chemically equilibrated.
In Fig.~\ref{fig:2} we show the profile of this configuration, as
mass-enclosed vs radius. Numerical simulations of the merger process
(e.g., \cite{Sekiguchi:2011mc}) show that, if the mass is not too
large, the merger remnant can survive longer than 10 ms (due to its rapid differential rotation)
before collapsing into a black-hole.
For the EoS we are using, a direct collapse will not occur for the
common case of the merger of two 1.3 $M_\odot$ stars, even neglecting
the additional stabilizing effect due to the stiffening of the EoS
\cite{Bauswein:2013jpa,Bauswein:2015,Bauswein:2015vxa}
 \footnote{The interesting scenario, in which at least one of the two
   stars is already a quark star, will be
   discussed in a future paper. The presence of stiff quark matter
   already inside the merger constituents will likely result in the
   merger remnant remaining stable up to the maximum mass allowed for
   a non-rotating quark star \cite{Kurkela:2009gj,Drago:2013fsa}.}.

After the conversion of the inner region to quark matter, what
follows is a process of much slower burning which, being no longer
accelerated by hydrodynamical instabilities, typically lasts a few
tens of seconds \cite{Drago:2015fpa}
\footnote{Here we use for $a_0$ of
  Ref.\cite{Drago:2015fpa} the central value $a_0=0.5$.}. The entire
star has converted to quark matter only after this slower burning
front has reached the remnant surface. We will show that
during this phase, no relativistic outflow - and hence no prompt GRB
emission - is expected from the merger remnant, similarly to what
happens in lGRBs. This is because in proto-magnetar models
the maximum achievable Lorentz factor of the flow is given by
$\Gamma_{\rm max}\sim\dot{E}/ \dot{M} c^2$, where $\dot{E}\sim B^2 R^6
(2\pi/P)^4 /3 c^3$ is the magnetic Poynting flux, $R$ is the stellar
radius, and $\dot{M}$ is the mass loss rate due to neutrino heating
\cite{Metzger:2010pp}. As long as the star maintains an external layer
of baryons, nucleons can be ablated from its surface by thermal
neutrinos with energies of a few MeV.

\begin{figure}
\vskip 0.5cm
\begin{centering}
\epsfig{file=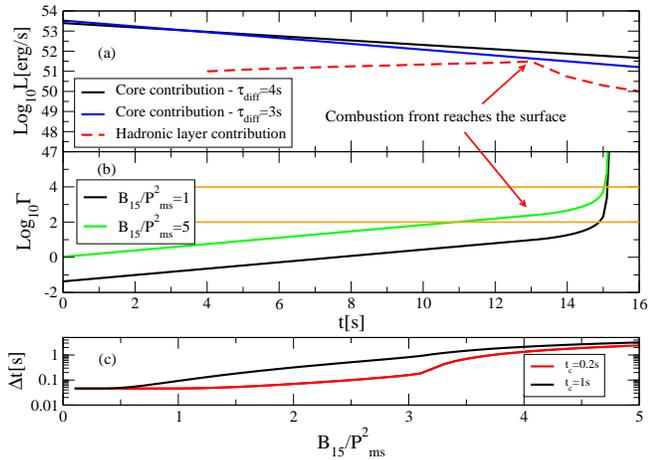,height=8.5cm,width=6cm,angle=-90}
\caption{{\it Panel (a)}: total neutrino luminosity. Solid lines
  correspond to the luminosity associated with the rapid burning of
  the central area (and two different values for the diffusion
  time). Dashed line the neutrino luminosity of the slow combustion of
  the external layer of the star. {\it Panel (b)}: Maximum bulk
  Lorentz factor of the magnetar jet, $\Gamma_{\rm max}$, as a
  function of time, shown for two values of $B/P^2$, where $B$ is the
  magnetic dipole and $P$ the rotation period. The two horizontal
  lines bracket the range of values of $\Gamma_{\rm max}$ required to
  produce GRB prompt emission according to conventional models. Here
  and in panel (a) the arrows indicate the time $t_0\sim$13 s at which
  the conversion of the remnant into a quark star is completed.  {\it
    Panel (c)}: duration of the prompt emission of the sGRB as a
  function of $B/P^2$, shown for two values of the time needed for baryon cessation $t_c$
(see text).}
\label{fig:3}
\end{centering}
\end{figure}

The evolution of $\dot{M}$ is quite complicated. 
During the first tenth of a second it reaches values as large as
$\dot{M} \sim 10^{-3}M_\odot s^{-1}$ \cite{Dessart:2008zd}.
In the following few seconds, the baryon flow is associated with 
the generation of protons via $\beta$-decay in the
cooling process.
In this way the remnant atmosphere becomes
progressively more proton rich, similar to the evolution of a
proto-neutron star after a supernova explosion. In our simple analysis
we borrow from the existing literature the result that $\dot{M}$
remains very large for a few seconds \cite{Metzger:2014ila} and we
assume that it can be approximated better and better with the formula
used in the case of a proto-neutron star after a supernova
explosion. In that case $\dot{M}$ is approximately given by
\cite{Qian:1996xt}:
\begin{equation}
\dot{M} \sim 1.2\times 10^{-9} C^{5/3}
L_{\overline{\nu}_e,51}^{5/3}\epsilon^{10/3}_{\overline{\nu}_e,\mathrm{MeV}} M^{-2}_{1.4}
R^{5/3}_{6} M_\odot s^{-1} \, ,
\end{equation}
where $L_{\overline{\nu}_e,51}$ is the electron anti-neutrino luminosity in units of
$10^{51}$ erg, $\epsilon_{\overline{\nu}_e,\mathrm{MeV}}$ is their energy in MeV, 
$M_{1.4}$ is the neutron star mass in units of 1.4 $M_\odot$, $R_{6}$ is the radius of the
star in units of 10$^{6}$ cm, and $C \sim 2$ is a correction factor to account for additional channels of neutrino heating \cite{Qian:1996xt}.  
The energy of neutrinos from the merger remnant is typically $\approx$ 10 MeV \cite{Perego:2014fma}.

The crucial ingredient in the calculation of $\dot{M}$, and hence
$\Gamma_{\rm max}$, is the neutrino luminosity. This has been
evaluated in \cite{Pagliara:2013tza}, accounting only for the heat
deposited during the rapid burning of the central region, while
\cite{Drago:2015fpa} also evaluates the emission associated with the
prolonged burning of the external layer. The contributions to the
neutrino luminosity from the initial phase of prompt burning in the
core, $L_\nu^{c}$, can be approximated in a simple way by introducing
the neutrino diffusion time $\tau_{\rm diff}$. Following
Ref. \cite{Pagliara:2013tza}:
\begin{equation}
L_\nu^{c}\sim Q/\tau_{\mathrm{diff}} \,\mathrm{e}^{-t/\tau_{\mathrm{diff}}}\, ,
\end{equation}
where $Q\sim (2-3)\times 10^{53}$ erg is the total heat deposited by
quark deconfinement during the rapid burning phase and $\tau\sim
2(3)$s for a star of mass 1.4(1.8) $M_\odot$, respectively. We employ
a similar formula in the merger case, but accounting for the larger
amount of heat deposited, $Q\sim 10^{54}$ erg (also due to the
gravitational potential energy before the merger and in part to the
use of a different equation of state), and $\tau_{\mathrm{diff}}\sim
3-4$ s, the latter estimated following Ref.\cite{Perego:2014fma}
(their eq.~6).

Fig.~\ref{fig:3} shows that, while the quark star is still forming,
the neutrino luminosity is very large and it corresponds to a mass
loss rate of $\approx 10^{-4} M_\odot s^{-1}$. Therefore the Lorentz
factor does not reach high enough values to produce the GRB prompt
emission. This stage mirrors the early evolution of the proto-magnetar
in lGRB, where no relativistic jet is created during the first
$\sim 10$ s after core bounce due to the high baryon load. 
In the case of lGRBs, after that phase the baryon load slowly reduces
and a GRB lasting a few tens seconds is produced. Notice that
in the case of lGRB, the mass of the proto-magnetar and its initial temperature
are significantly smaller and quark deconfinement need not to take place.
By contrast, in the merger case, the quark conversion is unavoidable and when the front reaches
the stellar surface baryonic ablation ceases. To zeroth order,
therefore, the prompt emission from the rotating magnetized merger
remnant is suppressed at all epochs: the mass loss rate is too large
prior to quark conversion, or too low after the conversion.  In
neither case can a prolonged relativistic outflow of the appropriate
Lorentz factor form.  In this zero-order approximation, the maximum
Lorentz factor $\Gamma_{\rm max}$ jumps from values of the order of
unity to, virtually, infinity.

Such a sudden jump in the outflow's Lorentz factor is clearly not
physical: what is missing is a description of the period over
which the most external layer of the star is converted into quarks.
Even if baryon loading were to cease abruptly, a minimum time would be
required to clear the jet of baryons, which we estimate to be
$t_\mathrm{d}\sim 0.01$ s as the dynamical timescale near the base of
the wind (Ref.~\cite{Vlasov+14}, Fig.~9).  However, there is a potentially more important effect that
delays the time for baryon cessation. Since the star is
rapidly rotating near centrifugal break-up, its shape is deformed into
an ellipsoid with an equatorial radius $R_{\mathrm{eq}}$ larger than
its polar radius $R_\mathrm{p}$. For a soft EoS, such as that we
employ for the hadronic phase, we expect
$R_{\mathrm{eq}}/R_\mathrm{p}\sim 1.2-1.4$ for a rotation rate of
$\sim 1$ kHz \cite{Bejger:2006hn}.  Using the results of
Ref.\cite{Drago:2015fpa}, we estimate that the burning front will
reach the pole and the equator at times $t_\mathrm{\rm p}$ and
$t_\mathrm{\rm eq}\approx (1.2-1.4)t_\mathrm{\rm p} $, respectively.
Since $t_p\sim (10-20)$ s, the quark conversion of the star will move from
pole to equator over a characteristic timescale of $\Delta t \sim
t_{\rm eq} - t_{\rm p} \sim $ a few seconds.

However, in fact baryon mass loss from the strongly magnetized remnant
is confined to a relatively narrow range of latitudes near the axis of
the magnetic dipole, which is likely to be aligned with the rotation
axis.  The latitudinal extent of this `open zone' of the magnetosphere
is given by $\theta_{\rm open} \approx \left(R/2R_{\rm L}\right)
\approx 0.1R_{6}(P/\rm 2 ms)^{-1}$, where $R_{\rm L} = 2\pi P c$ is
the light cylinder radius.  Thus, for typical values of $P \sim 2$ ms,
we expect the true timescale for baryon cessation to be given by
$t_c \approx \theta_{\rm open}(\Delta t \sim t_{\rm eq} - t_{\rm p}) \sim$ a few
$0.1$ s, comparable to the duration of sGRBs.

Fig.~\ref{fig:3} shows our results for
the duration of the sGRB prompt emission, which we indeed find to be
of the right order of magnitude. sGRBs of the longest duration may
start even during the final seconds of the baryon emission, before
deconfinement reaches the surface (as occurs if $\dot E$ is very
large), while the shortest duration are instead regulated by
$t_\mathrm{d}$.  Interestingly, we predict a strong correlation
between the sGRB duration and its luminosity (which is $\propto
B^2/P^4$), which is indeed observed \cite{Shahmoradi:2014ira}.

Finally, in our model it is possible to have precursor signals: since
the inner engine is already active during the first ten seconds, some high energy emission can originate from the
jet before the main event starts. Precursors have indeed been
observed from sGRBs \cite{Troja:2010zm}.

%\bibliography{references}
%\bibliographystyle{apsrev}

\end{document}